\documentclass[aps,twocolumn,showpacs]{revtex4}
\usepackage{graphicx}
\begin{document}
\title{Transport, optical properties and quantum ratchet effects for 
quantum dots and molecules coupled to Luttinger liquids.}
\author{A. Komnik$^{1}$ and A. O. Gogolin$^{2}$}

\affiliation{$^{1}$Physikalisches Institut, Albert--Ludwigs--Universit\"at,
D--79104 Freiburg, Germany \\
$^{2}$Department of Mathematics, Imperial College London, 180 Queen's Gate,
London SW7 2BZ, United Kingdom}

\date{\today}

\begin{abstract}
We present non-perturbative solutions for  
multi-level quantum dot structures
coupled to interacting one-dimensional electrodes out of equilibrium. 
At a special correlation strength the Hamiltonian can be mapped to the Kondo
problem which possesses a solvable Toulouse point, where all
conductance and noise properties can be calculated exactly. Special attention
is paid to the fully asymmetric setup when each dot level is 
coupled to only one of the leads and the electron 
transport through the structure is accompanied by photon absorption (emission).
A relation between the optical
spectra and the energy dependent current noise power is established. 
Experimental implications of the results, specifically for the
Fano factor, the ratchet current, and field emission via localised states, 
are discussed. In particular, we predict that the peak in the
ratchet current as function of the irradiation frequency splits
up in two due to correlation effects.

\end{abstract}

\maketitle
\section{Introduction}
Manufacturing of micro- and nano-electronic circuitry based on single 
molecules represents one prospective way to achieve further miniaturisation 
as well as efficiency improvement of electronic devices. 
First successful
attempts of contacting single molecules have been reported in a number of
recent experimental works \cite{reed,c60,weber,h2contact}. One possible
mechanism for the electron transport through them is tunnelling on and off the
molecular orbitals (MOs). The smaller the molecule the larger is the energy
distance between the MOs so that in some cases the transport occurs through
only one electron level even at room temperatures. Hence, the adequate 
physical description of such systems coincides with that of the single-state
quantum dot (QD): a fermionic level coupled to metallic 
electrodes (we shall also call them `leads' or `contacts'). 

If one aims at small device dimensions one has to go for 
one dimensional (1D) electrodes. 
Promising candidates for such
wirings are the carbon nanotubes in their single-wall version
(SWNTs) \cite{dekkerPT,avouris}. However, truly one-dimensional electron
systems cannot be described by the Fermi liquid (FL) model. No matter how weak
the fermion interactions are, they cannot be taken into account perturbatively.
It is well known that in the low energy sector the interacting 1D fermions
constitute a universality class of Luttinger liquids (LL), which display
a completely different physics than the conventional FLs \cite{book}. 
As a consequence, the electronic degrees of freedom in sufficiently thin
SWNTs are also described by a generalisation of the LL model
\cite{gogeg,kanebalentsfisher}. This has also been confirmed experimentally
\cite{bockrath}. 

In the most of existing experiments the contacts between the molecule and 
the leads are quite weak. The optimal operation of the future nano-electronic
devices is, however, expected in setups with small contact resistances.
That can be achieved only in systems where the current-mediating MOs
undergo strong hybridisation with the valence bands of the
electrodes. In the case when two MOs with different symmetries
couple to different leads electronic transport can only take place via
emission or absorption of photons. 
Transport resulting from such `optical' coupling can be distinguished from the 
background transport by its dependence on the laser irradiation frequency. 
In spite of vast amount of contributions dealing with QD's
transport and optical properties (see, for example, 
\cite{schoeller}, and references therein), systems
with good contacts did not receive much attention. 
Moreover, to the best of our knowledge, 
the issue of interacting electrodes has been discussed only in 
resonant tunnelling context
\cite{kanefisher,furusakidot,NG,restunn,gornyi,huegle}. 
In this paper we begin to close these gaps and
present non-perturbative solutions for multi-level dots
contacted by interacting 1D electrodes.
We wish to clarify here that these type of correlations
are different from `on-dot' couplings (Coulomb or Hubbard terms).
The latter coupling gives rise to the Kondo type phenomena
\cite{kouwenhoven}. In this paper we neglect the on-dot
couplings and effectively deal with spinless electrons.
This restricts the validity of our results to the temperatures
above the Kondo temperature $T_K$ or to the case when a polarising
magnetic field is present. Such an approach appears to be
justified for SWNTs, which display strong LL correlations in a
wide temperature range from about 5 K to about 100 K
\cite{bockrath}, typical $T_K$'s being much smaller.
The interplay of on-dot and `in-lead' correlation is
of theoretical interest but remains outside the scope of
this paper.

The outline of the paper is as follows. 

In Section \ref{noise},
we start with a review of the simplest realisation: a single-state dot coupled
to 1D electrodes. These results have been recently announced in
our Letter \cite{restunn}. 
Contrary to \cite{restunn} we work in the Green's function 
formalism, which is more suited to access the noise properties. Since the
transport in such a setup is completely understood both in the
resonant case as well as in the off resonant case, our primary
goal in the Section \ref{noise} is the study of the Fano factor, 
which is a
ratio of the zero frequency noise and the transport current, as a function
of bias voltage. For the sake of completeness we derive all
equations needed to access the full energy dependence of the noise power
spectrum.   

In Section \ref{multiple},
we go over to a two-state QD with both levels coupled to 1D leads.
Since the non-interacting situation is
more or less trivial we restrict our considerations only to the interacting
system. It turns out, that, as in the case of the single-state QD, the 
Hamiltonian can again be brought to a quadratic form in terms of new
fermions for a special interaction strength 
\cite{restunn}. In this representation the non-linear transport
as well as zero frequency noise properties can be expressed via the
transmission coefficient even in case of an additional tunnelling term 
between the dot levels. 
Our non-perturbative approach allows then to study 
all resonant tunnelling effects (known for single-state setups \cite{NG})
in this situation. 

In Section \ref{ratchetsection} we investigate
transport in a similar structure where every level is only coupled to one of
the electrodes and where the dominant transport mechanism is the photon-assisted
tunnelling between the levels. That situation corresponds to 
coupling of two MOs of different symmetries discussed above. It turns out
that a finite current can flow even without any applied voltage. That makes
such a system one of the simplest realisation of the so-called `quantum ratchet'
effect \cite{reimann,linke,lehmann}. 
We concentrate on the analysis of the `ratchet'
current as a function of the radiation frequency. 
Some general results, including an important
relation between the current noise power and the absorption and emission
spectra are discussed in Section \ref{preliminary}. A treatment of  
non-interacting systems follows in Section \ref{NIratchet}. 
Transport through a dot coupled to LL electrodes is then analysed in 
Section \ref{Iratchet}. 

A short summary of results (Section \ref{summary}) concludes the paper.
We stress again that though we
model the molecule--electrodes coupling by means of tunnelling
Hamiltonians, all our results are non-perturbative in 
tunnelling amplitudes contrary to the bulk of 
existing studies \cite{schoeller}.

\section{Transport through a single-state quantum dot}      
                   \label{noise}
                   
First we briefly review the method of \cite{restunn} and then
discuss the noise properties.                   

\subsection{Scattering states solution and duality}

We model the system by the following Hamiltonian (we ignore the spin 
throughout the paper):
\begin{eqnarray}                     \label{H0}
 H = H_K + H_t + H_C \, ,
\end{eqnarray}
where $H_K$ is the kinetic part,
\begin{eqnarray} \nonumber
H_K = \Delta d^\dag d + \sum_{i=R,L} H_0[\psi_i] \, ,
\end{eqnarray}
describing the electrons in the leads $H_0[\psi_i]$,
and the resonant level $\Delta$, the corresponding electron
operators being $d^\dag,d$. The dot can be populated from
either of the two leads ($i=R,L$) via electron tunnelling with amplitudes
$\gamma_i$,
\begin{eqnarray}          \nonumber
 H_t = \sum_i  \gamma_i[ d^\dag \psi_i(0) + \mbox{h.c.}] \, .
\end{eqnarray}
In (\ref{H0}), $H_C$ describes the electrostatic
Coulomb interaction between the leads and the dot,
\begin{eqnarray} \nonumber
 H_C =  \lambda_C d^\dag d \, \sum_i \, \psi_i^\dag(0) \psi_i(0) \, .
\end{eqnarray}
This interaction is a new ingredient we have introduced, absent in the related
studies \cite{NG} and \cite{kanefisher}. It does not, however, affect the
universality as we shall show later.

The contacting electrodes are supposed to be one-dimensional
half-infinite electron systems. We model them
by chiral fermions living in an infinite system: the negative half-axis
then describes the particles moving towards the boundary,
while the positive half-axis
carries electrons moving away from the end of the system.
In the bosonic representation $H_0[\psi_i]$ are diagonal even in presence
of interactions (for a recent review see e.g. \cite{book};
we set the renormalised Fermi velocity $v=v_F/g=1$,
the bare velocity being $v_F$):
\begin{eqnarray}                 \label{Hi}
 H_0[\psi_i] = (4 \pi)^{-1} \int \, dx \, [\partial_x \phi_i(x)]^2.
\end{eqnarray}
Here the phase fields $\phi_i(x)$ describe the slow varying
spatial component of the electron density (plasmons),
\begin{eqnarray} \nonumber
\psi^\dag_i(x) \psi_i(x) = \partial_x \phi_i(x)/2 \pi \sqrt{g} \, .
\end{eqnarray}
The electron field operator at the boundary is given by \footnote{Strictly
speaking $\psi(x=0)=0$, so we
assume that the tunnelling takes place at the second last site of the
corresponding lattice model, at $x=\pm a_0$. Also, we ignore the Klein factors
as they can be absorbed into Eqs.(\ref{newfermions})
disappearing from the analysis.},
\begin{eqnarray}                           \label{psioperator}
 \psi_i(0) = e^{i \phi_i(0)/\sqrt{g}}/\sqrt{2 \pi a_0} \, ,
\end{eqnarray} 
where $a_0$ is the lattice constant of the underlying lattice model.
Here $g$ is the conventional LL parameter (coupling constant) connected to the
bare interaction strength $U$ via $g=(1+U/\pi v_F)^{-1/2}$
\cite{book,kanefisher}. In the chiral formulation the bias voltage amounts
to a difference in the densities of the incoming particles in both channels
far away from the constriction \cite{eggergrabert}. The current is then
proportional to the difference between the densities of incoming 
and outgoing particles within each channel.

To the best of our knowledge, Hamiltonian (\ref{H0}) cannot be solved exactly
even in the $g=1$ case as long as $\lambda_C$ remains finite. However, after
a transformation of $d^\dag$ and $d$ operators to the spin representation
of the form
\begin{eqnarray} \nonumber
 \left\{  \begin{array}{l}
                  S_x = (d^\dag + d)/2,
                  \\
                  S_y = - i (d^\dag - d)/2,
                  \\
                  S_z = d^\dag d - 1/2,
                  \end{array} \right.
           \; , \;
\end{eqnarray}
one immediately observes that  the $\lambda_C$ term is
analogous to the $S_z$--spin density coupling in the Kondo problem.
The latter is known to be explicitly solvable at a
particular value of the longitudinal coupling:
the Toulouse limit (see e.g. \cite{book}).
Let us perform a similar calculation.
As a first step we introduce new symmetric and antisymmetric
fields
\begin{eqnarray}                  \label{symasymphi}
\phi_\pm = (\phi_L \pm \phi_R)/\sqrt{2} \, ,
\end{eqnarray} 
which still fulfill the bosonic commutation relations. Then we apply the 
transformation $H'=U^\dag H U$ with \cite{emerykivelson}
\begin{eqnarray}   \nonumber
 U=\exp( i S_z \phi_+(0)/ \sqrt{2 g}) \, ,
\end{eqnarray}
which changes the kinetic and the Coulomb coupling parts of the full
Hamiltonian to [we drop a constant contribution proportional to $S_z
\delta(x)$ that does not affect the transport]
\begin{eqnarray}      \nonumber
 H_K'+ H_C' = H_K +
 (\lambda_C/\pi \sqrt{2g} - \sqrt{2/g}) S_z \partial_x
 \phi_+(0) \, , \nonumber
\end{eqnarray}
and the tunnelling part (terms containing $\gamma_i$) to
\begin{eqnarray}
 H_t' &=& (2 \pi a_0)^{-1/2} \Big[ S_+ (\gamma_L e^{i \phi_-/\sqrt{2g}} +
 \gamma_R e^{-i \phi_-/\sqrt{2 g}})
 \nonumber \\ \nonumber
 &+& (\gamma_L e^{-i \phi_-/\sqrt{2g}} +
 \gamma_R e^{i \phi_-/\sqrt{2g}}) S_- \Big] \, ,
\end{eqnarray}
where $S_{\pm}= S_x \pm i S_y =d^\dag,d$. At the point $g=1/2$ one can
re-fermionise the problem by defining new operators
\begin{eqnarray}              \label{newfermions}
\psi_\pm = e^{i \phi_\pm}/\sqrt{2 \pi a_0} \, ,
\end{eqnarray}
which  fulfill standard fermionic commutation relations. With the help of
the particle density operator $\psi^\dag_\pm \psi_\pm =
\partial_x \phi_\pm/2 \pi$ we can immediately write down the refermionised
Hamiltonian,
\begin{eqnarray}                   \label{Htransformed}
 H &=& H_0[\psi_\pm] + (\lambda_C - 2 \pi) 2S_z \psi_+^\dag \psi_+ + \Delta S_z
 \nonumber \\
 &+& S_+ (\gamma_L \psi_- + \gamma_R \psi^\dag_-) + (\gamma_L \psi^\dag_- +
 \gamma_R \psi_-) S_- \, .
\end{eqnarray}
In the case of the symmetric coupling $\gamma_L=\gamma_R$ this Hamiltonian
is similar to that of the two-channel Kondo problem and, at the Toulouse point
$\lambda_C = 2 \pi$, can be solved exactly (out of equilibrium)
using the method of Ref.\cite{SH}.
The novel ingredient in the following analysis is the
extension to the asymmetric case.
To take advantage of the Toulouse point we set the
Coulomb coupling amplitude to $2\pi$ in what follows.
This not only removes the four fermion interaction
but decouples the `$\pm$' channels making the `$+$' channel free
(i.e. decoupled from the dot variables).

At the Toulouse point our Hamiltonian describes free fermions
which scatter at the origin. These new fermions are 
related to the physical electrons in a highly non-local way. 
As the relations between the particle densities are 
still linear, 
in order to access the transport properties it is sufficient to 
calculate the energy dependent transmission coefficient 
$1-T(\omega)$ of the new fermions [$T(\omega)$ being the transmission
coefficient of the physical ones]. 
The non-linear $I-V$ characteristics is then given by
\begin{eqnarray}                         \label{IVdefinition}
 I(V) = G_0 \int \, d \omega \, T(\omega) [n_F(\omega-V) - n_F(\omega)] \, ,
\end{eqnarray}
where $n_F$ denotes the Fermi distribution function 
and $G_0 = e^2/h$ is the conductance quantum. 
The easiest way to identify $T(\omega)$ is the equation of motion 
method. We calculated $T(\omega)$ in Ref. \cite{restunn}, 
it is given by (we measure all energies in units of $\Gamma =
\gamma_L^2 + \gamma_R^2$): 
\begin{eqnarray}                       \label{1-D}
 &&T(\omega) \\ \nonumber 
 &=&\frac{ 4 \gamma^2 E^2}{(E^2 + \beta_+^2)(E^2 + \beta_-^2) + 2
 \gamma^2 (E^2 + \beta_- \beta_+) + \gamma^4} \, ,
\end{eqnarray}
where 
\begin{eqnarray}       \nonumber        
 E&=&\Delta^2-\omega^2 \, , \\ \nonumber 
 \beta_\pm &=& [(1-2\alpha)\Delta \pm \omega]/2 \, , \\ \nonumber 
 \gamma &=& \omega \sqrt{\alpha(1-\alpha)} \, ,
\end{eqnarray}
and $\alpha=\gamma_L^2/(\gamma_L^2 + \gamma_R^2)$ is the asymmetry parameter.

Using expressions (\ref{IVdefinition}) and (\ref{1-D}) one can access all
conductance properties of the system. 
We shall not discuss them again (see Ref.\cite{restunn}) 
but concentrate instead on the duality property.
In the simplest case of the symmetric model on-resonance ($\alpha=1/2$,
$\Delta=0$) we obtain
\begin{eqnarray}            \label{Tomegaresonance}
 T(\omega) = (1+\omega^2)^{-1} \, . 
\end{eqnarray}
As a consequence, the temperature  dependent differential conductance 
at zero bias (which is the most relevant quantity from the 
experimental point of view) amounts to 
\begin{equation}                      \nonumber
 G_{\Delta=0}(T)/G_0=\frac{1}{2\pi T}
 \psi' \left( \frac{1}{2}+\frac{1}{2\pi T}\right) \, ,
\end{equation}
where $\psi$ denotes the $\psi$--function. Comparing this result with the
conductance $G_{1/2}$ through a single scatterer in an LL with $g=1/2$, 
given in Ref. \cite{kanefisher}, we find that (similar relations are 
valid for the non-linear $I-V$'s)
\begin{eqnarray}            \label{g12} 
 G_{\Delta=0}(T)/G_0=1-G_{1/2}(T)/G_0 \, ,
\end{eqnarray}
where $T$ in $G_{1/2}(T)$ is measured in units of the
backscattering strength.
According to the duality hypothesis the strong coupling fixed point at $g$
corresponds to the weak coupling one at $1/g$ and vice versa, 
leading to the relationship of the conductances of the form
\cite{kanefisher,weiss1,fls},  
\begin{eqnarray}            \label{g2}
 G_{2}(T)/G_0=1-G_{1/2}(T)/G_0 \, .
\end{eqnarray}
Therefore, Eqs.(\ref{g12}) and (\ref{g2}) suggest that tunnelling between
two LLs with $g=1/2$ via a resonant level is equivalent to direct 
tunnelling between two LLs with $g=2$. 

In fact, we can demonstrate this
equivalence on the Hamiltonian level. We start with (\ref{Htransformed})
and in order to simplify things introduce new real (Majorana)
fermions $a,b$ and $\zeta,\eta$ according to 
\begin{eqnarray}                      \label{MajoranaDef}
 d=(a+ib)/\sqrt{2} \, \, , \, \, \psi_- = (\zeta+i\eta)/\sqrt{2} \, .
\end{eqnarray}
In this language the Hamiltonian acquires the form
\begin{eqnarray}                       \label{MajoranaHam}
 H = H_0[{\zeta,\eta,a,b}] + i \gamma_- a \eta(0) + i \gamma_+ b \zeta(0) \, ,
\end{eqnarray}
where $\gamma_\pm = \gamma_L \pm \gamma_R$ and the unperturbed part is 
defined by
\begin{eqnarray} 
 H_0[{\zeta,\eta,a,b}] &=& i \Delta a b + i \int\, dx \, \Big[ \eta(x) 
 \partial_x \eta(x) + \zeta(x) \partial_x \zeta(x) \nonumber \\
 &+& V \zeta(x) \eta(x) \Big] \, . \nonumber
\end{eqnarray}
For future reference, we give the current operator in this 
representation:
\begin{eqnarray}                     \nonumber
 J = - \frac{i}{2} \gamma_- a \zeta(0) - \frac{i}{2} \gamma_+ b \eta(0) \,.
\end{eqnarray}

On the other hand, the Hamiltonian for the direct tunnelling 
between two LLs in terms of the physical fermions is 
\begin{eqnarray}                     \nonumber
 H = \sum_{i=R,L} H_0[\psi_i] + \gamma \left( \psi_L^\dag \psi_R + \psi_R^\dag
 \psi_L \right) \, ,
\end{eqnarray}
containing the free part $H_0$, which again describes two 
half-infinite LLs, and tunnelling between them with the amplitude
$\gamma$. We bosonize the above Hamiltonian as in the
previous Section using rules (\ref{Hi}) and (\ref{psioperator}), and 
obtain 
\begin{eqnarray} \nonumber
 H = \sum_i H_0[\phi_{i}] + \frac{\gamma}{2 \pi a_0} \left[ e^{i
 (\phi_L-\phi_R)/\sqrt{g}} + e^{-i(\phi_L-\phi_R)/\sqrt{g}} \right] \, .
\end{eqnarray}
Introducing new fermions (\ref{newfermions}) at the point $g=2$
(we drop the `+' channel again since it is free)
\begin{eqnarray} \nonumber
 H = H_0[\psi_-] + \frac{\gamma}{\sqrt{2 \pi a_0}} (
 \psi_-^\dag + \psi_-) \, .
\end{eqnarray}
Next we take advantage of a trick from
Ref.\cite{matveevtrick} and perform a substitution $\Psi = (d-d^\dag) \psi_-$,
where $d$ is some local fermionic operator \emph{not related} in any way
to the dot operator of the previous Section. 
Obviously, such transformation does not change either 
the commutation relations or the normalisation of the 
operators. Then
\begin{eqnarray} \nonumber
 H = H_0[\Psi] +  \frac{\gamma}{\sqrt{2 \pi a_0}}(c - c^\dag)(\Psi^\dag +
 \Psi) \, .
\end{eqnarray}
The last step is obvious: one introduces the Majorana components
according to (\ref{MajoranaDef}). This results in 
\begin{eqnarray} \nonumber
 H = H_0[\zeta,\eta] + i \gamma \sqrt{\frac{2}{\pi a_0}} b \zeta(0) \, ,
\end{eqnarray}
which is precisely the Hamiltonian (\ref{MajoranaHam}) of a resonant setup
($\gamma_-=0$, $\Delta=0$ and, of course $V=0$) up
to the redefinition $\gamma_+ = \gamma \sqrt{2/\pi a_0}$. 
We have checked that the bias voltage and
the current operator of the $g=2$ LL problem and
the resonant tunnelling system transform correctly. 


\subsection{Green's functions solution and noise properties} \label{noisefano}

Although the transport properties can easily be accessed by means of the
scattering formalism as shown in Ref.\cite{restunn}, 
it is not immediately clear (see below though) how the 
information about the finite frequency noise can be extracted
from the transmission coefficient. A more appropriate method to calculate
the fluctuations is the Green's functions (GFs) method in its non-equilibrium
(Keldysh) formulation. For further reference we now define all possible
non-equilibrium GFs. Let $\mu,\nu$ stand for either of 
the electrode Majoranas $\zeta$ or $\eta$ (taken at $x=0$) and 
$f,h$ stand for either the dot level operators, $a$ or $b$.
Then define
\begin{eqnarray}                        
 D_{f h}^{ij} (t-t') = - i \langle T_C f(t) h(t') \rangle \, , \nonumber \\
 G_{\mu \nu}^{ij} (t-t') = - i \langle T_C \mu(t) \nu(t') \rangle \, , 
 \nonumber \\
 G_{\mu f}^{ij}(t-t') = - i \langle T_C \mu(t) f(t') \rangle \, , 
 \nonumber \\
 G_{f \mu}^{ij}(t-t') = - i \langle T_C f(t) \mu(t') \rangle \, , 
 \nonumber 
\end{eqnarray}
where $T_C$ is the time ordering operator along the Keldysh 
contour $C$, which 
consists of the forward $C_-$ and the backward $C_+$ paths. The times $t,t'$
belong to the paths $C_{i,j}$, respectively. Sometimes we shall omit
the Keldysh indices $i,j$ in what follows adopting matrix notation for
the Keldysh GFs. 
In the above definitions we assumed the
system to be in a steady state so that all GFs are translationally invariant
in the time domain and therefore depend only on the time differences. 
There is an obvious relation 
$G_{\mu f}^{ij}(t-t') = - G_{f \mu}^{ij}(t'-t)$. 
However, working with two definitions possesses advantages as we shall see
later.

It is, in fact, not difficult to calculate
the zero order GFs, when $\gamma_i=0$. 
For the electrode Majoranas we obtain
\begin{eqnarray}                     \label{zetazetanull}
 G^{(0)}_{\zeta \zeta}(\omega) =  G^{(0)}_{\eta \eta}(\omega) =
 \frac{i}{2}\left(  \begin{array}{cc}
                  H(\omega) & H(\omega)+1
                  \\
                  H(\omega)-1 & H(\omega)
                  \end{array} \right)
           \; , \;
\end{eqnarray}
where $H(\omega)$ contains the information about the Fermi distribution
functions $n_F$ of the original electrons in the leads,
$H(\omega)=n_F(\omega+V)-n_F(-\omega+V)$. Obviously, the cross correlations
$G^{(0)}_{\zeta \eta}$ exist only as long as the applied voltage is finite,
\begin{eqnarray}                     \label{zetaetanull}
 G^{(0)}_{\zeta \eta}(\omega) = - G^{(0)}_{\eta \zeta}(\omega) =
 \frac{1}{2} F(\omega) \left(  \begin{array}{cc}
                  1 & 1
                  \\
                  1 & 1
                  \end{array} \right)
           \; . \;
\end{eqnarray} 
This fact is reflected by the function $F(\omega) = n_F(\omega +
V)-n_F(\omega-V)$ vanishing as $V\to 0$. 
As a consequence of the special forms of
(\ref{zetazetanull}) and (\ref{zetaetanull}) the retarded 
and the advanced components are fairly simple: 
$G^{(0) R,A}_{\mu \nu}=0$ while $G^{(0) R,A}_{\mu \mu}= \mp i/2$. 
The $a-b$ subsystem being in equilibrium
makes the calculation of the corresponding GFs even simpler. The
result is 
\begin{eqnarray}
 D^{(0) --(++)}_{f f} &=& \pm \frac{1}{2} \sum_{p=\pm}
 [\omega+p(\Delta-i\delta)]^{-1} \, , \nonumber \\ \nonumber
 D^{(0) -+(+-)}_{f f} &=& \pm i \pi \delta(\omega+\Delta) \, , \\ \nonumber
 D^{(0) --(++)}_{a b} &=& \mp \frac{i}{2} \sum_{p=\pm}
 [\Delta-p(\omega+i\delta)]^{-1} \, , \nonumber \\ \nonumber
 D^{(0) -+(+-)}_{a b} &=& \mp \pi \delta(\omega+\Delta) \, .
\end{eqnarray}
Using these zero order GFs as a starting point we can calculate any
correlation function exactly because of the quadratic form 
of the Hamiltonian. Our goal is to calculate the average of the 
current operator $I(V)=\langle J \rangle$
and the power spectrum of current fluctuations (we also shall call this
quantity noise spectrum), which is defined as 
\begin{eqnarray}
 P(\Omega) = \int d \, t \, e^{i\Omega t} \, \left[ \langle J(t)
 J(0)\rangle - \langle J(0) \rangle^2 \right] \, .
\end{eqnarray}
The averages are calculated using the $S$ matrix for the coupling of
dot and electrode Majoranas, $\langle \dots \rangle = \langle \dots S 
\rangle_0$, which is
\begin{eqnarray}  \nonumber
 S = T_C \exp\left( - \int_C \, d\tau \, \gamma_- a(\tau) \eta(\tau) +
 \gamma_+ b(\tau) \zeta(\tau) \right) \, .
\end{eqnarray}
Expanding in powers of $\gamma_i$ one can derive for the current the
analog of the Meir-Wingreen formula \cite{meirwingreen},
\begin{eqnarray}                    \label{meirwingreen}
 I(V) = \frac{i}{8 \pi} \int \, d \omega \, F(\omega) [ \gamma_+^2
 D^A_{bb}(\omega) - \gamma_-^2 D^A_{aa}(\omega)] \, .
\end{eqnarray} 
We choose to split the noise spectrum into two contributions, 
$P(\Omega) = P_\parallel(\Omega) + P_\perp(\Omega)$, 
where both quantities can be expressed in terms of off-diagonal 
(containing different fermion species) GFs, 
\begin{widetext}
\begin{eqnarray}                         \label{noisepar} 
 P_\parallel(\Omega) = \frac{1}{4} \int \, d \omega \, 
 \gamma_+^2 \Big[ G^{+-}_{b\eta}(\omega) G^{+-}_{\eta b}(\Omega-\omega) 
 &-& D^{+-}_{bb}(\omega) G^{+-}_{\eta \eta} (\Omega-\omega)\Big] 
 \\ \label{noiseperp}
  + \gamma_-^2 \Big[  G^{+-}_{a\zeta}
 (\omega) G^{+-}_{\zeta a}(\Omega-\omega) &-& D^{+-}_{aa}(\omega) G^{+-}_{\zeta
 \zeta} (\Omega-\omega)\Big] \, , \\ 
 P_\perp(\Omega) = \frac{\gamma_- \gamma_+}{4} \int \, d\omega \, \Big[
 G^{+-}_{a \eta}(\omega)  G^{+-}_{\zeta b}(\Omega-\omega) &+& G^{+-}_{b
 \zeta}(\omega)  G^{+-}_{\eta a}(\Omega-\omega) - 2 D^{+-}_{ab}(\omega)
 G^{+-}_{\zeta \eta}(\Omega-\omega) \Big] \, . 
\end{eqnarray}
\end{widetext}

By means of the $S$ matrix expansion one can reduce some of the 
off-diagonal GFs to the diagonal ones. In particular, for the functions
entering the parallel part of the noise spectrum we obtain 
(we omit the energy variable $\omega$), 
\begin{eqnarray}                        \label{betaexact}
 G^{+-}_{b \eta} = i \frac{\gamma_+}{2} F D^R_{bb} - i \gamma_- 
 (D^{(0) +-}_{ab} G^A_{\eta \eta} + D^{(0) R}_{ab} G^{+-}_{\eta \eta})
\end{eqnarray}
Similar expressions can be derived for $G^{+-}_{\eta b}$, $G^{+-}_{a \zeta}$
and $G^{+-}_{\zeta a}$. 
GFs entering the $P_\perp$ part of the noise have 
somewhat different structure, 
\begin{eqnarray} \nonumber
 G^{+-}_{a \eta} = -i \frac{\gamma_+}{2} F D^R_{ab} - i \gamma_- (D^{+-}_{aa}
 G^{(0) A}_{\eta \eta} + G^{(0) +-}_{\eta \eta} D^{A}_{aa}) \, , 
\end{eqnarray} 
where the remaining GFs, $G^{+-}_{a \eta}$, $G^{+-}_{b \zeta}$ and 
$G^{+-}_{\zeta b}$ are similar. The remaining 
off-diagonal GFs cannot be reduced to the diagonal ones.
They should rather be found as solutions of a chain of Dyson equations. 
For the $a-b$ subsystem one
obtains the following system of equations (to simplify notation we ignore
here the Keldysh indices): 
\begin{widetext} 
\begin{eqnarray}                     \label{dysonab}
 D_{ab} &=& D^{(0)}_{ab} + \gamma_+^2 D^{(0)}_{ab} G^{(0)}_{\zeta \zeta} D_{bb}
                       + \gamma_-^2 D^{(0)}_{aa} G_{\eta \eta} D^{(0)}_{ab}
                       + \gamma_+ \gamma_-( D^{(0)}_{ab} G^{(0)}_{\zeta \eta}
                       D_{ab} + D^{(0)}_{aa} G^{(0)}_{\zeta \eta} D_{bb})
 \\  \label{dysonbb}
 D_{bb} &=& D^{(0)}_{bb} + \gamma_+^2 D^{(0)}_{bb} G^{(0)}_{\zeta \zeta}D_{bb}
                       + \gamma_-^2 D^{(0)}_{ba} G^{(0)}_{\eta \eta} D_{ab}
                       + \gamma_- \gamma_+ (D^{(0)}_{ba} G^{(0)}_{\eta \zeta}
                       D_{bb} + D_{bb} D^{(0)}_{\zeta \eta} D^{(0)}_{ab})
\end{eqnarray}
In the same way one can derive the corresponding equations for the electrode
Majorana GFs: 
\begin{eqnarray}                        \label{dysonzetaeta}
 G_{\zeta \eta} &=& G^{(0)}_{\zeta \eta} + \gamma_+^2 G^{(0)}_{\zeta \zeta}
 D^{(0)}_{bb} G_{\zeta \eta} + \gamma_-^2 G^{(0)}_{\zeta \eta} D^{(0)}_{aa}
 G_{\eta \eta} - \gamma_+ \gamma_- ( G^{(0)}_{\zeta \zeta} D^{(0)}_{ab} G_{\eta
 \eta} - G^{(0)}_{\zeta \eta} D^{(0)}_{ab} G_{\zeta \eta}) \,, \\
 \label{dysonetaeta}
 G_{\eta \eta} &=&  G^{(0)}_{\eta \eta} + \gamma_+^2 G^{(0)}_{\eta \zeta}
 D^{(0)}_{bb} G_{\zeta \eta} + \gamma_-^2 G^{(0)}_{\eta \eta} D^{(0)}_{aa}
 G_{\eta \eta} + \gamma_+ \gamma_- ( G^{(0)}_{\eta \zeta} D^{(0)}_{ba} G_{\eta
 \eta} + G^{(0)}_{\eta \eta} D^{(0)}_{ab} G_{\zeta \eta}).
\end{eqnarray}
\end{widetext}
In the simplest symmetric case $\gamma_-=0$ we obtain from Eq.(\ref{dysonbb})
for the advanced dot level GF 
\begin{eqnarray} \nonumber
 D^{A}_{bb} = D^{(0) A}_{bb}/(1+\gamma_+^2 D^{(0) A} G^{(0) A}_{\zeta \zeta})
 \, .
\end{eqnarray}
Plugging this result into the expression for the current 
(\ref{meirwingreen}) 
results in Eq.(\ref{IVdefinition}) 
with all the energy variables measured in units of
$\Gamma=\gamma_+^2/4$ we identify 
\begin{eqnarray}                         \label{symoffresTomega}
T(\omega) = \omega^2/[ (\omega^2-\Delta^2)^2 + \omega^2] \, , 
\end{eqnarray}
as precisely the transmission coefficient 
(\ref{1-D}) at $\alpha=1/2$ found previously by 
means of the equations of motion method. 

As for the noise spectrum, only the parallel component survives. 
Eq.(\ref{dysonbb}) has the following solution: 
\begin{eqnarray} \nonumber
 D^{+-}_{bb} = \frac{ D^{(0) +-}_{bb} - \Gamma |D^{(0) R}_{bb}|^2 G^{(0)
 +-}_{\zeta \zeta}}{| 1 + \Gamma D^{(0) R}_{bb} G^{(0) R}_{\zeta \zeta}|^2} \,
 ,
\end{eqnarray} 
while its electrode counterpart,
\begin{eqnarray} \nonumber
 G^{+-}_{\eta \eta} = G^{(0) +-}_{\eta \eta} + 2 \Gamma F D^{(0) A}_{bb}
 G^A_{\zeta \eta} \, ,
\end{eqnarray}
is related to the cross--correlation $G^A_{\zeta \eta}$. 
The latter is a solution to one of the equations in 
(\ref{dysonzetaeta}), 
\begin{eqnarray}  \nonumber
G^{A}_{\zeta \eta} = G^{(0) A}_{\zeta \eta}/(1+\gamma_+^2 D^{(0) A} G^{(0) A}_{\zeta \eta})
 \, ,
\end{eqnarray}
which is zero. Therefore we have $G^{+-}_{\eta \eta} = G^{(0) +-}_{\eta
\eta}$. The rest of the needed GFs can be read off Eq.(\ref{betaexact}),
$G^{+-}_{b \eta} = i \gamma_+ F D^R_{bb}/2$, and $G^{+-}_{\eta b} = -i 
\gamma_+ F D^A_{bb}/2$. Collecting all terms in (\ref{noisepar}) we obtain
the following result:
\begin{widetext}
\begin{eqnarray}                          \label{generalPsym}
 P_{sym}(\Omega) = e G_0 \int \, d \omega \, \Big\{ -  \frac{F(\omega)
 F(\Omega - \omega) \omega(\Omega-\omega)}{(\omega^2 - \Delta^2 - i
 \omega)[ (\Omega-\omega)^2 - \Delta^2 + i (\Omega-\omega)]} 
 + \frac{[H(\omega)-1][H(\Omega-\omega)-1] \omega^2}{|\omega^2 - \Delta^2 - i
 \omega|^2} \Big\} \, .
\end{eqnarray}
\end{widetext}
The same formula has been obtained by Schiller and Hershfield 
(SH) in Ref.\cite{SH} in the context of the non-equilibrium 
Kondo problem, where the  magnetic field strength plays the role
of our level energy $\Delta$.

In the following we do not repeat the results for the noise spectrum 
which are already contained in Ref.\cite{SH} but rather concentrate 
on the aspects which are specific to resonant tunnelling, 
in particular on the asymmetric case and on the calculation of the 
Fano factor not covered by SH. 

In the limit of zero frequency we 
find the noise spectrum to be given by the formula  
\begin{eqnarray}               \label{zeroomeganoise}
 P_{sym}(0) &=& e G_0 \int \, d \omega \, T(\omega)[1-T(\omega)] 
 \nonumber \\ &\times&
 [n_F(\omega-V) - n_F(\omega)]  \, . 
\end{eqnarray}
identical to the one derived for non-interacting electrons
\cite{martin}. This is somewhat surprising. The reason must be 
that we map the original Hamiltonian onto a free one, where
the current carrying excitations are again of the fermionic nature. 
That is why in order to access $P(0)$ even in the asymmetric case
we do not have to solve the above Dyson equations but can simply 
use formula (\ref{1-D}) for the transmission coefficient.

Let us pause here to mention that,
in non-interacting resonant tunnelling systems at zero
temperature, the Fano factor $\nu_V = P(0)/e I(V)$ is 
suppressed in comparison to the Schottky value $\nu=1$ by the 
factor $(\Gamma_L^2 + \Gamma_R^2)/(\Gamma_L+\Gamma_R)^2$ 
at high voltages $V\gg \Gamma$
and by the factor $(\Gamma_L-\Gamma_R)^2/(\Gamma_L+\Gamma_R)^2$ in the
opposite limit $V\ll \Gamma$ \cite{chenting}. 
As previously $\Gamma_{L(R)}$ denote the dimensionless conductances of the 
left(right) contact. The suppression is maximal, 
$\nu_\infty=1/2$ and $\nu_0=0$, in the symmetric 
case $\Gamma_R=\Gamma_L$. 
According to Ref.\cite{braggio}, as soon as we deal 
with an LL system, the Fano factor is expected to keep 
its maximal value no matter how strong the asymmetry is.
 
At zero temperature and on--resonance we
find using (\ref{Tomegaresonance}) that the on-resonance shot 
noise is (we also recover the correct pre-factors):
\begin{eqnarray}                   \nonumber
 P_{sym}(0) = \frac{e G_0}{2} \left[ \tan^{-1} V - V/(1+V^2) \right] \, , 
\end{eqnarray}
Taking into account the formula for current, $I(V) = G_0 \tan^{-1}V$, 
one can read off the Fano factor, which has the following 
limiting forms: 
\begin{eqnarray}                             \nonumber
\nu_{V\rightarrow 0} &=& \frac{1}{3} V^2 + O(V^4) \, , \nonumber \\
\nu_\infty &=& \frac{1}{2} \, .
\end{eqnarray} 

In the general asymmetric situation the transmission coefficient 
takes the form 
\begin{eqnarray} \nonumber
 T(\omega) = \frac{4 \alpha  (1-\alpha) 
 \omega^2}{\left[ \omega^2 + 1/4 + 
\alpha(1-\alpha)\right]^2
 - \alpha(1-\alpha)} \, .
\end{eqnarray}
The evaluation of the Fano factor now yields 
\begin{eqnarray}                            \label{nu1}
 \nu_0 &=& 1 \, , \nonumber \\
 \nu_\infty &=& 2 \alpha^2 -2 \alpha+1 = \frac{\Gamma_L^2 +
 \Gamma_R^2}{(\Gamma_L + \Gamma_R)^2} \, .
\end{eqnarray}
At high voltages, $V\gg \Gamma$, we recover the noise suppression of 
the non-interacting case. On the contrary, our value for $\nu_0$ is 
in apparent contradiction to the results of Ref. \cite{braggio}. 
The reason for that discrepancy is quite simple. Ref. \cite{braggio} assumes
the sequential tunnelling process to be the dominant transport mechanism. As
was pointed out in \cite{kanefisher,furusakidot} this is indeed the case for
not too low temperatures for arbitrary $g$ and even at $T=0$ (that is in our 
situation) as long as the interactions are strong enough, for $0<g<1/2$. These
conditions are obviously not compatible with our assumptions. 

\begin{figure}
\vspace*{0.5cm}
\includegraphics[scale=0.30]{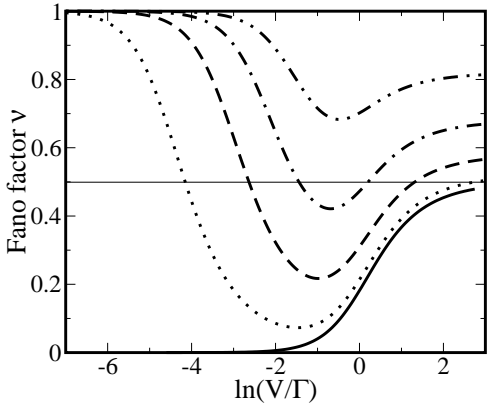}
\caption[]{\label{fanofactoromeganode} 
 The Fano factor as a function of the bias voltage for different 
 asymmetry values ($\alpha=0.5$,$0.4$,$0.3$,$0.2$,$0.1$ from the bottom
 curve upwards).}
\vspace*{-0.5cm}
\end{figure}

It is, in fact, not difficult to access the full crossover behaviour 
of the Fano factor, see Fig.\ref{fanofactoromeganode}.
The most striking feature of the full plot is the presence of 
the local minimum
at $V^*$ as long as the system is kept either symmetric or out of
resonance. 
$V^*$ can be shown to be the solution of $\nu_{V^*} = T(V^*)$, so
that it gives precisely the point at which the 
transmission coefficient crosses $\nu_V$. 
Similar local minima have also been
found in \cite{braggio}.

In the case of a symmetric system off-resonance,
where the transmission coefficient is given by 
Eq.(\ref{symoffresTomega}), the
emerging picture is fully consistent with (\ref{nu1}). In
the small bias limit the Fano factor approaches unity, $\nu_0 = 1$, 
whereas in
the limit of high voltages we again recover the universal non-interacting
noise suppression as $\nu_\infty = 1/2$. The $\nu_V$ behaviour is
qualitatively the same as in the asymmetric case, see
Fig. \ref{fanofactoromeganode}, including the minimum at intermediate
voltage. The asymptotic value at $\nu_{\infty}=1/2$ is 
now universal for all curves. 

We performed a study of the general case $\alpha\neq 1/2$ and 
$\Delta \neq 0$
as well. The limiting behaviour of $\nu_V$ turns out to be 
determined solely 
by the asymmetry parameter $\alpha$ and is completely independent of the
detuning $\Delta$. For $\Delta \gg 1$ the position $V^*$ of the 
intermediate 
minimum is asymmetry independent and coincides with $\Delta$. 

Contrary to the zero frequency noise the evaluation of $P(\Omega)$ 
spectrum at finite $\Omega$ and, in the general case of an asymmetric system
off-resonance, requires knowledge of the full transmission amplitude 
matrix, 
as one already can see from (\ref{generalPsym}) 
\cite{blanter}. The latter is not at all 
the same in both fermionic 
representations of the problem, so that we have to solve the full 
set of 
Dyson equations (\ref{dysonab})-(\ref{dysonetaeta}). 
The full solution is rather lengthy. It does not appear to
reveal qualitatively new features as compared to what is already 
known \cite{SH} as the effects of asymmetry and finite $\Delta$ 
do not compete but rather enhance each other. We shall therefore
conclude the discussion of the noise spectrum at this point.

\section{The two--state quantum dot}             \label{multiple}
\begin{figure}
\vspace*{0.5cm}
\includegraphics[scale=0.30]{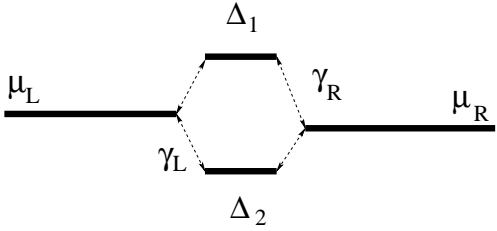}
\caption[]{\label{nonratchet} 
 Quantum dot with two levels.}
\vspace*{-0.5cm}
\end{figure}

In the spirit of the previous Section we model the double 
QD by two fermionic 
levels with energies $\Delta_{1,2}$, which are coupled to LL leads, see
Fig. \ref{nonratchet}. 
As long as the levels do not interact with each other in any 
other way then by
tunnelling, the whole treatment including the solution of the equations of
motion can be performed for an arbitrary number of levels. Throughout this
Section we are not interested in noise power spectra so that we 
concentrate 
only on the conductance properties of the system, which are most easily
accessed by means of the equation of motion method. 
The Hamiltonian of the system is
still assumed to be of the form (\ref{H0}) with following changes:
  
(i) The kinetic part describes two (or more) levels instead 
of only one, 
\begin{eqnarray}  \nonumber
H_K = \sum_{i=1,2}\Delta_i d_i^\dag d_i + \sum_{i=R,L} H_0[\psi_i]. 
\end{eqnarray}

(ii) The tunnelling amplitudes are the same for both levels. It is, in fact,
not difficult to solve the problem with arbitrary amplitudes. 
This, however, does not induce new physics, so we restrict our 
solution to this special case:
\begin{eqnarray} \nonumber
H_t = \sum_{i} \sum_{j=R,L} 
\gamma_j[ d_i^\dag \psi_j(0) + \mbox{h.c.}] 
\, .
\end{eqnarray}

(iii) The strength of the electrostatic Coulomb interaction is 
also assumed to be the same for both levels, 
\begin{eqnarray} \nonumber
 H_C =  \lambda_C \sum_{i} d_i^\dag d_i \, \sum_j \, \psi_j^\dag(0)
 \psi_j(0) \, . 
\end{eqnarray}

(iv) There is an additional term in the Hamiltonian that is responsible for
the tunnelling processes between the dot levels \footnote{If one 
considers the levels $\Delta_{1,2}$ to be
the molecular orbitals \emph{after} the contacting process there is 
no need to include the tunnelling. 
However, we assume the dot level wave functions to be
those of a free molecule prior to contacting. 
In that case there could be some
finite overlap, and hence tunnelling, between them.}:
\begin{eqnarray} \nonumber
 H_W = W ( d_1^\dag d_2 + \mbox{h.c.}) \, .
\end{eqnarray}

To describe the electronic degrees of freedom in the electrodes we use the
same formalism as in the previous Section, see 
Eqs. (\ref{Hi})-(\ref{psioperator}). 
As in the case of a single level we can introduce symmetric
and anti-symmetric
components (\ref{symasymphi}) and spin representations for the
level operators (they acquire an index $i=1,2$) and apply a slightly different
transformation to the overall $H$, defined by
\begin{eqnarray}                       \label{EKtrafo2}
 U=\exp \left( i \sum_i S_i^z \phi_+/ \sqrt{2 g}\right) \, .
\end{eqnarray}
This transformation changes the kinetic and the Coulomb 
coupling parts of the full 
Hamiltonian to (we again drop a constant contribution)  
\begin{eqnarray} \nonumber
 H_K'+ H_C' = H_K + 
 (\lambda_C/\pi \sqrt{2g} - \sqrt{2/g}) \sum_i S_i^z \partial_x
 \phi_+(0) \, , \nonumber 
\end{eqnarray}
and the tunnelling part (terms containing $\gamma_i$) to
\begin{eqnarray}
 H_t' &=& (2 \pi a_0)^{-1/2} \Big[ \sum_i S_i^+ (\gamma_L e^{i
 \phi_-/\sqrt{2g}} + \gamma_R e^{-i \phi_-/\sqrt{2 g}}) 
 \nonumber \\ \nonumber
 &+& (\gamma_L e^{-i \phi_-/\sqrt{2g}} +
 \gamma_R e^{i \phi_-/\sqrt{2g}}) S_i^- \Big] \, ,
\end{eqnarray}
where $S^{\pm}= S_i^x \pm i S_i^y =d_i^\dag,d_i$. The intra-dot
tunnelling term $H_W$ is invariant under this transformation.
The refermionisation can again be performed using the definitions
(\ref{newfermions}) and the resulting Hamiltonian differs from that in
(\ref{Htransformed}) only by the sums over both spins,
\begin{eqnarray}                   \label{Htransformed2}
 H &=& H_0[\psi_\pm] + \sum_i (\lambda_C - 2 \pi) 2  S_i^z \psi_+^\dag 
 \psi_+ + \Delta_i S_i^z
 \nonumber \\
 &+& S_i^+ (\gamma_L \psi_- + \gamma_R \psi^\dag_-) + (\gamma_L \psi^\dag_- +
 \gamma_R \psi_-) S_i^- \, .
\end{eqnarray}
In what follows we concentrate on the Toulouse point when
$\lambda_C = 2 \pi$, where the `$\pm$' channels decouple and when the
Hamiltonian acquires a very convenient quadratic form. 

In order to calculate the non-linear $I(V)$ we employ the method of
Ref.\cite{restunn}, which results in Eq.(\ref{IVdefinition}) with 
some modified $T(\omega)$. 
As in the case of the single-level dot, the easiest way to find 
the transmission coefficient is the equations of motion method. 
Since we have 
two types of operators: for the 
electrons of the `$-$' channel and for the dot levels (we go back 
to the original $d_i^\dag, d_i$ operators), 
we need two types of equations of motion, 
\begin{eqnarray}                    \label{EoMs}
 i \partial_t \psi_-(x) &=& -i \partial_x \psi_-(x) + \sum_i \delta(x)(\gamma_L
 d_i -
 \gamma_R d_i^\dag) \, , \nonumber \\
 i \partial_t d_i &=& \Delta_i d_i + W d_{-i} + \gamma_L \psi_-(0) + 
\gamma_R \psi_-^\dag(0) \, . 
\end{eqnarray}
Integrating the first one around $x=0$ with respect to $x$ from $-\epsilon$
to $\epsilon$ and then sending $\epsilon$ to zero we obtain
\begin{eqnarray}            \label{afterepsint}
 i [\psi_-(0^+)-\psi_-(0^-)] = \sum_i \gamma_L d_i - 
 \gamma_R d_i^\dag \, 
\end{eqnarray}
where $0^\pm$ denotes positive (negative) infinitesimal.
We define new operators 
\begin{eqnarray}                 \label{xitrdef}
 Y &=& \prod_{i} (i\partial_t - \Delta_i) - W^2 \, , \nonumber \\
 L &=& 2 ( i \partial_t + W ) - \sum_i \Delta_i \; , \nonumber \\
 Z_\pm &=& \partial_t^2 + i \sum_i \Delta_i \partial_t \pm (\Delta_0 \Delta_1
 - W^2) \; .  
\end{eqnarray}
By acting with $|Y|^2$ on both sides of Eq.(\ref{afterepsint}) 
and using the
last two equations (\ref{EoMs}) we can eliminate the dot operators. We obtain
as a result
\begin{eqnarray}                          \label{polnoye}
 |Y|^2 [\psi(0^+) - \psi(0^-)] = (\gamma_L^2 Z_+ L + \gamma_R^2 Z_- L^*)
 \psi(0) \nonumber \\ 
 + \gamma_L \gamma_R ( Z_+ L + Z_- L^*) \psi^\dag(0) \, .
\end{eqnarray}
Now we can insert into this relation the momentum decomposition of the field
operator $\psi_-$ 
\begin{eqnarray}                    \label{partialdecomposition}
 \psi_-(x,t) = \int \, \frac{d k}{2 \pi} e^{i k (t-x)} 
\left\{  \begin{array}{l}
                  a_k \, \, \, \mbox{for} \, x<0
                  \\
                  b_k \, \, \, \mbox{for} \, x>0
                  \end{array} \right. \, .
\end{eqnarray}
Because the dispersion relation is linear, $\omega = v k = k$, 
we can use $\omega$ as the momentum variable as well as the 
energy variable. 
Inserting Eq.(\ref{partialdecomposition}) into Eq.(\ref{polnoye})
and using $\psi_-(0) = [\psi_-(0^+) + \psi_-(0^-)]/2$ results in
a following equation, which form 
(up to a redefinition of constant factors) 
is independent of the number of levels, 
\begin{eqnarray}                    \label{finalequation}
E(b_\omega - a_\omega) = -i \beta_+ (a_\omega + b_\omega) + i \gamma 
(a^\dag_{-\omega} + b^\dag_{-\omega})
\, .
\end{eqnarray}
We introduced the following objects,   
\begin{eqnarray}                       \label{Egammabetadefinitions}
 E &=& \prod_{p=\pm} \left[\prod_{i} (\omega + p \Delta_i)-W^2 \right] 
 \, , \nonumber \\ 
 \gamma &=& \omega \sqrt{\alpha(1-\alpha)} \Big[ \sum_{j} \Delta_j^2 -
 2 \omega^2 \nonumber \\
 &-& 2 W \left( \sum_i \Delta_i - W \right) \Big] \, , \nonumber \\
 \beta_\pm &=& (1/2-\alpha) \Big[ \sum_{j} \Delta_j \prod_{m\neq j} 
 (\Delta_m^2 - \omega^2) - 2 W \omega^2 \nonumber \\ 
 &-& W^2 \sum_i \Delta_i 
 \Big] \pm \gamma/2 \sqrt{\alpha(1-\alpha)} \, . 
\end{eqnarray}
Formally Eq. (\ref{finalequation}) has exactly the same form as in the
single-level case. Therefore the resulting transmission coefficient
is still given by formula (\ref{1-D}) with modified 
constants contained in (\ref{Egammabetadefinitions}). 

The temperature behaviour of $G$ at the maxima does not differ considerably
from that for one single level, which has already been studied in
Ref.\cite{restunn}. 
\begin{figure}
\vspace*{0.5cm}
\includegraphics[scale=0.38]{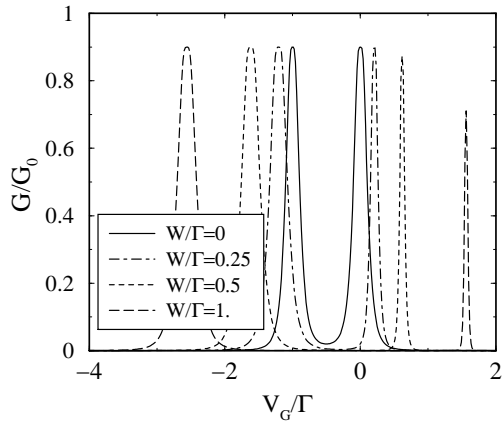}
\caption[]{\label{symform} 
Linear differential conductance of a symmetric setup as a function 
of the gate voltage $V_G$ at the temperature $T=0.01\Gamma$ 
at different 
values of the intra-dot tunnelling amplitude. The bare dot level energies
are $\Delta_0=-\Gamma$ and $\Delta_1=0$.}
\vspace*{-0.5cm}
\end{figure}
\begin{figure}
\vspace*{0.5cm}
\includegraphics[scale=0.38]{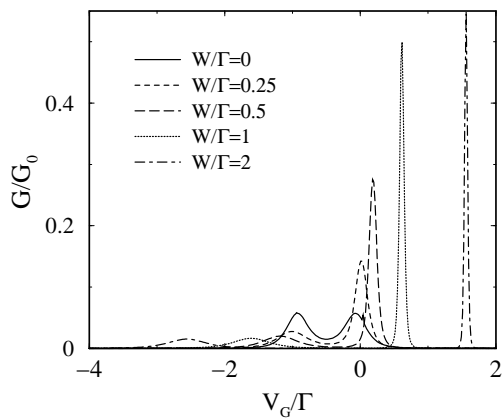}
\caption[]{\label{asymform} 
The same plot as in Fig.\ref{symform} but for an asymmetric system
with $\alpha=0.25$.}
\vspace*{-0.5cm}
\end{figure}
The conductance $G(T)$ in the valley between the peaks
turns out to be very well described by a superposition of two peaks of 
the single level problem. Moreover, 
the presence of the $W$ tunnelling process
does not affect the conductance either, 
only leading to some suppression of
current at low temperatures $T\ll \Gamma_{R,L}$. 

The issue of the peak shape is far more interesting, see
Figs. \ref{symform},\ref{asymform}. In the asymmetric 
system the upper peak tends to sharpen and to increase in height with 
growing tunnelling amplitude whereas the lower peak suffers the 
opposite fate, see Fig. \ref{asymform}. 
This can be understood in the picture where
the Hamiltonian is diagonalised with respect to $W$ tunnelling. 
The corresponding
transformation is given by a rotation in the two-dimensional space 
of operators $d_{1,2}$ to the new ones $\widetilde{d}_{1,2}$ 
via the rotation matrix $R(\alpha)$, where the angle $\alpha$ is given by
\begin{eqnarray} \nonumber
 \tan 2\alpha = \frac{2 W}{\Delta_2-\Delta_1} \, .
\end{eqnarray}
Then the energies of the two new levels are
\begin{eqnarray} \nonumber
 \widetilde{\Delta}_{1,2} = \frac{\Delta_1+\Delta_2}{2} \pm \sqrt{
 \left(\frac{\Delta_1-\Delta_2}{2}\right)^2 + W^2} \, .
\end{eqnarray}
The couplings to individual levels are subject to change as a 
consequence of the operator sum transformation
\begin{eqnarray} 
 \nonumber
 d_1+d_2 = \sqrt{2} \cos(\alpha+\pi/4) \widetilde{d}_1 + \sqrt{2}
 \cos(\alpha-\pi/4) \widetilde{d}_2 \, ,
\end{eqnarray}
which makes the coupling of the upper level decrease with growing 
tunnelling amplitude $W$ when $\alpha$ tends to $\pi/4$. 
The coupling of the lower level, on the contrary, increases. 
Notice that such renormalisation occurs even in the
non-interacting systems. The physical reason is that the conductance via
tunnelling through the lower level is enhanced because 
of the additional depopulation process which transfers 
electrons to the upper level. The
non-trivial interaction effect in our setup is the increasing height of the
upper level. The reason is that due to smaller $\Gamma_{R,L}$ the 
upper level is at effectively higher temperature, which, in turn, means that
the conductance is higher in the asymmetric case. 
For the same reason the height of the lower peak diminishes.  

The coupling symmetry is not affected by the 
diagonalisation transformation, 
so that the resonant conductance is increasing monotonically all the way
to zero temperature. That is why in this case the upper peak amplitude is
lower, see Fig. \ref{symform}.

\section{Quantum dot interacting with photons: A quantum ratchet setup}
\label{ratchetsection}

\subsection{Preliminary considerations}       \label{preliminary}
   
\begin{figure}
\vspace*{0.5cm}
\includegraphics[scale=0.30]{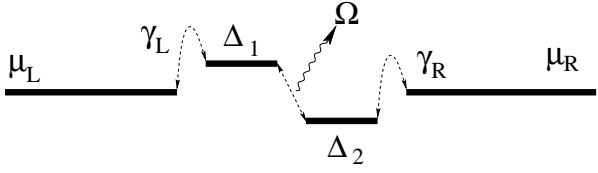}
\caption[]{\label{ratchet} 
The quantum ratchet setup.}
\vspace*{-0.5cm}
\end{figure}

Now we slightly change the setup. 
The kinetic and Coulomb coupling terms 
$H_K+H_C$ remain the same while the coupling to the leads becomes completely
asymmetric: the level $\Delta_1$ is coupled only to the left lead and 
$\Delta_2$ only to the right lead (see Fig.\ref{ratchet}), 
\begin{eqnarray}                   \label{newHt}
 H_t =  \gamma_L d_1^\dag \psi_L(0) + \gamma_R d_2^\dag \psi_R(0) + \mbox{h.c.}
\end{eqnarray}
We assume that the localised levels possess 
different symmetries so that a 
direct tunnelling between them is forbidden (nevertheless, the Coulomb
coupling, being free of selection rules, is still symmetric) 
while hopping with simultaneous 
emission or absorption of a photon with energy $\Omega$ is possible. 
Then the coupling of the levels can be written as
\begin{eqnarray} \nonumber
 H_W = W \left( e^{i \Omega t} d_1^\dag d_2 + e^{-i \Omega t} d_2^\dag d_1
 \right) \, ,  
\end{eqnarray}
where $W$ denotes the coupling amplitude. We dropped the photon operators
since we assume that the dot (or molecule) is subject to intensive laser
radiation (so that there is \emph{always} 
a phonon which can be absorbed). 

Although the full Hamiltonian is now explicitly time dependent 
it still can 
be reduced to a system in a steady state via the gauge transformation 
\begin{eqnarray}               \label{gaugetrafo}
 d_1 &\rightarrow& d_1 e^{i \Omega t/2} \, , \nonumber \\
 d_2 &\rightarrow& d_2 e^{-i \Omega t/2} \, .
\end{eqnarray}
Thereby the energy levels of the dot states as well as the chemical potentials
in the leads are shifted to 
\begin{eqnarray} \nonumber
 \Delta_{1,2} &\rightarrow& \Delta_{1,2} \mp \Omega/2 \, , \nonumber \\ 
 \mu_{R,L} &\rightarrow& \mu_{R,L} \mp \Omega/2 \, . \nonumber
\end{eqnarray}
Thus, the coupling to the radiation effectively results in a finite 
bias voltage. As a consequence, a finite current can flow
without any real applied voltage at all. 
This is the \emph{quantum ratchet effect}
which has recently been intensively discussed in
Ref.\cite{reimann,linke,lehmann} in various non-interacting setups with mostly
weak couplings to the leads.  

The physical explanation of the effect in the original picture prior to 
the gauge transformation is quite simple. We start with a system where 
the two dot levels possess two different energies, 
$\Delta_2<\Delta_1$ as in Fig.\ref{ratchet}. 
The population probability of the lower of level ($\Delta_2$) is
higher than that of its counterpart $\Delta_1$, so that the photon 
absorption is the dominant process transferring the electrons to 
$\Delta_1$. They can relax either to the left lead or back to $\Delta_2$.
However, if the hybridisation of the dot levels is larger than the 
electromagnetic coupling, the dominant  relaxation 
process is tunnelling into the left lead. 
This leads to a non-zero net current through the system.  

The quantities we are interested in are again the full current $I(V)$ through
the system, the current noise power $P(\omega)$ and the light 
absorption (emission)
$A(\omega)[E(\omega)]$ spectra. The first quantity can be defined e.g. 
via the expectation value of the particle flow between the left lead 
and $\Delta_1$, 
\begin{eqnarray}                          \label{toknormal}
 \hat{I} = -i \gamma_L [ d_1^\dag \psi_L(0) - \psi_L^\dag(0) 
d_1 ] \,.
\end{eqnarray}
In the language of non-equilibrium Keldysh diagram technique 
it is given by the following expression \cite{keldysh,LLX,QE}, 
\begin{eqnarray}                \label{tokformula} 
 I &=& \frac{\gamma_L^2}{2 \pi} \int \, d\omega \, \Big[ G^{(0)+-}_L(\omega)
 D^{-+}_1(\omega) \nonumber \\
 &-& G^{(0)-+}_L(\omega)
 D^{+-}_1(\omega) \Big] \, , 
\end{eqnarray}
where $-+(+-)$ indices stand for lesser and greater Keldysh GFs. 
For the left (right) lead electrons they are denoted by 
$G^{ij}_{L,R}(\omega)$ while for the dot electrons by 
$D^{ij}_{1,2}(\omega)$. 
The additional superscript $(0)$ distinguishes the GFs in the absence of
tunnelling couplings. Alternatively one can define the current as the expectation
value of the tunnelling operator between the dot states, 
\begin{eqnarray}               \label{tokalternat}
 \hat{I}' = -i W (d^\dag_1 d_2 - d^\dag_2 d_1) \; .
\end{eqnarray} 
The third possible method to describe the transport is the transmission 
coefficient formalism.
Using definition (\ref{tokalternat}) we derive an expression for 
the noise power spectrum of the form:
\begin{eqnarray}
 P(\omega) &=& W^2 \int \, dt \, e^{i \omega t} \Big[ \langle d^\dag_1(t)
 d_2(t) d^\dag_2(0) d_1(0) \rangle \nonumber \\ \nonumber
 &+& \langle d^\dag_2(t)
 d_1(t) d^\dag_1(0) d_2(0) \rangle \Big] - 2 \pi \delta(\omega) I^2 \, .
\end{eqnarray}

The light absorption rate can be evaluated via 
the Golden Rule and is given by 
\begin{eqnarray} \nonumber
 A(\omega) = 2 \pi W^2 |\langle f | d^\dag_1 d_2 |0 \rangle|^2
 \delta(\omega-\Delta_1 +\Delta_2) \, ,
\end{eqnarray}
where the vacuum $|0\rangle$ is assumed to be the state with
the level $\Delta_1$ empty and $\Delta_2$ full. The final state $|f\rangle$
is just the opposite. It turns out that the Fourier transform of 
the function 
\begin{eqnarray}                   \label{Sat}
 S^A(t) = - i \langle d^\dag_2(t) d_1(t) d^\dag_1(0) d_2(0) \rangle
\end{eqnarray}
is directly related to the absorption rate, 
\begin{eqnarray} \nonumber
 A(\omega) = i W^2 S^A(\omega) \, . 
\end{eqnarray}
Another interesting aspect of Eq.(\ref{Sat}) is the fact that it can be
written down in the form
\begin{eqnarray}                     \label{SA}
 S^{A}(\omega) = -\frac{i}{2 \pi} \int \, d\epsilon \, 
 D^{-+}_{2}(\epsilon) D^{+-}_{1}(\epsilon+\omega) \, .
\end{eqnarray}
This formula is exact as long as the GFs are calculated exactly and the 
level operators participate only in two-particle interaction
terms (such as $H_W$ or $H_t$). 

Similarly one can show that the emission rate $E(\omega)$ is 
proportional to a related function
\begin{eqnarray}                    \label{Eomegadef}
 E(\omega) = i W^2 S^E(\omega) \, , 
\end{eqnarray}
where 
\begin{eqnarray}                    \label{Etdef}
 S^E(t) = - i \langle d^\dag_1(t) d_2(t) d^\dag_2(0) d_1(0) \rangle \, ,
\end{eqnarray}
the Fourier transform having the same form as (\ref{SA}) up to the
exchange $1\leftrightarrow 2$.
Thus we establish a very convenient expression relating the 
total optical spectrum with the noise power of our system:
\begin{eqnarray}                     \label{Pviaspectrum}
 P(\omega) = A(\omega) + E(\omega) - 2 \pi \delta(\omega) I^2 \,
 . 
\end{eqnarray}
This relation has far reaching consequences for the field emission (FE)
physics \cite{plummer}. In the case when the chemical potential of
the right lead is sent to $\mu_R = - \infty$ the double dot setup describes
tunnelling into vacuum through a sequence of localised states, which is
nothing else than FE \cite{FEPRB}. 
Experimentally, it turns out that under certain conditions 
the FE from carbon nanotubes is accompanied by luminescence phenomena
\cite{french1}. One possible explanation has been offered in
Ref. \cite{french1}. It was suggested that localised levels
on the nanotube tip play the dominant role during the light
emission. Therefore, measuring the noise power spectrum along with the
luminescence spectrum during FE experiments and comparing them with the
prediction (\ref{Pviaspectrum}) would allow one to check 
the hypothesis put forward in Ref. \cite{french1}. 
However, we would like to postpone the detailed discussion of this very 
extensive issue to a later publication and rather concentrate here on the
ratchet effects. 
(We shall still give some general formulae for the 
noise power spectrum.) 

\subsection{The non-interacting case}          \label{NIratchet}

\begin{figure}
\vspace*{0.5cm}
\includegraphics[scale=0.40]{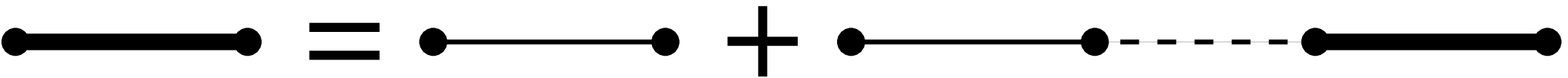}
\caption[]{\label{dyson1} 
Diagrammatic representation of the Dyson equation (\ref{dysoninkeldysh}).
Solid lines represent the GFs of the level $\Delta_1$ and 
the dashed lines those of $\Delta_2$. Thick lines are the exact GFs.
}
\vspace*{-0.5cm}
\end{figure}

We first discuss the case of non-interacting leads 
(and $\lambda_C=0$). Then the lead electron fields can
be integrated out exactly. As a result we are then left with zero-dimensional
problem of two fermionic levels. The corresponding Keldysh GFs at zero--order 
in tunnelling are given by the following matrices:
\begin{widetext}
\begin{eqnarray}                \label{matrixN1} 
  D_{1,2}^{(0)}(\omega) = \frac{1}{(\omega-\Delta_{1,2})^2 +
                  \Gamma_{L,R}^2}\left(  \begin{array}{cc} 
                  \omega-\Delta_{1,2}-i \Gamma_{L,R}\, \mbox{sgn}(\omega\mp
                  V/2) 
                  & i 2 \Gamma_{L,R} \Theta(-\omega \pm V/2)
                  \\ \\
                  -i 2 \Gamma_{L,R} \Theta(\omega \mp V/2) &
                 - \omega+\Delta_{1,2}-i \Gamma_{L,R}\, \mbox{sgn}(\omega\mp
                  V/2) 
                  \end{array} \right) \, .
\end{eqnarray}
\end{widetext}
The energies $\Delta_{1,2}$ are already renormalised by the gauge
transformation and $\Gamma_{R,L} = \rho_{R,L} \gamma_{R,L}^2$ where
$\rho_{R,L}$ are the density of states (DOS) in the corresponding lead. 
In the absence of an external bias voltage $V=\Omega$. 
The tunnelling term $H_W$ is
quadratic so that we can write down the exact Dyson equation 
for the full GFs of the dot levels. 
In terms of the Keldysh contour ordered GFs it can be
written as
\begin{eqnarray}                 \label{dysoninkeldysh}
 D_{1,2}(t-t') &=& D^{(0)}_{1,2}(t-t') + W^2 \int_C \, dt_1 \, dt_2 \,
 D^{(0)}_{1,2}(t-t_1)
 \nonumber \\ &\times&
 D^{(0)}_{2,1}(t_1-t_2) D_{1,2}(t_2-t') \, .
\end{eqnarray}
The corresponding diagrams are depicted in Fig.\ref{dyson1}. 
Disentangling the indices one obtains the following set of equations, 
\begin{eqnarray}                      \label{dysonsystem}
 D^{ij}_{1,2}(\omega) =  D^{(0)ij}_{1,2}(\omega) - \sum_{m=\pm} m
K^{im}_{1,2}(\omega) D^{mj}_{1,2}(\omega) \, , 
\end{eqnarray}
with the kernels 
\begin{eqnarray} \nonumber
 K^{ij}_{1,2}(\omega) = - W^2 \sum_{m=\pm} m D^{(0)im}_{1,2}(\omega) 
 D^{(0)mj}_{2,1}(\omega) \, .
\end{eqnarray}
Equation system (\ref{dysonsystem}) is linear and can easily be 
solved for all GFs. 
We need only the following ones (we drop the trivial energy argument
$\omega$):  
\begin{eqnarray}                  \label{Dsolution}
 D^{+-}_{1,2} &=& \frac{ D^{(0)--}_{1,2}K^{+-}_{1,2} + 
 [1-K^{--}_{1,2}] D^{(0)+-}_{1,2}}{\mbox{det} K_{1,2}} \\ \nonumber
 D^{-+}_{1,2} &=& \frac{ -D^{(0)++}_{1,2}K^{-+}_{1,2} + 
 [1+K^{++}_{1,2}] D^{(0)-+}_{1,2}}{\mbox{det} K_{1,2}} \, ,
\end{eqnarray} 
where 
\begin{eqnarray}
\mbox{det} K_{1,2} &=& (1-K^{--}_{1,2})(1+K^{++}_{1,2})
 +K^{-+}_{1,2}K^{+-}_{1,2} \nonumber \\ \nonumber
 &=& | 1 + K^R_{1,2}|^2
 \, .
\end{eqnarray}
Evaluation of the current with help of (\ref{tokformula}) and
(\ref{Dsolution}) yields at zero temperature the following result:
\begin{eqnarray}                    \label{noninttok}
 I = G_0 \int_{-V/2}^{V/2} \, d \omega \, T(\omega) \, ,
\end{eqnarray}
where the transmission coefficient is given by
\begin{widetext}
\begin{eqnarray}                      \label{nonintTomega}
 T(\omega) =  \frac{4 \alpha_L \alpha_R}{[(\omega-\Delta_1)^2 +
\alpha_L^2][(\omega-\Delta_2)^2 + \alpha_R^2] - 2 [
(\omega-\Delta_1)(\omega-\Delta_2) - \alpha_R \alpha_L] + 1} \, .
\end{eqnarray}
\end{widetext}
From now on all energy variables are normalised to $W$ and 
dimensionless, $\alpha_{R,L}= \Gamma_{R,L}/W$. 
\begin{figure}
\vspace*{0.5cm}
\includegraphics[scale=0.45]{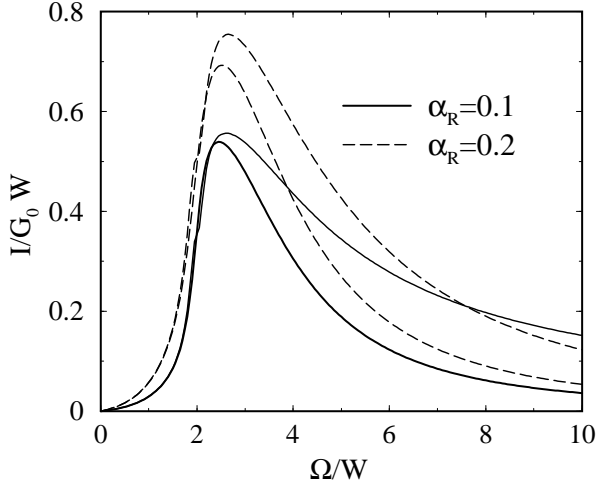}
\caption[]{\label{Tzeroratchet} 
Light induced current through a quantum ratchet system without the
applied voltage as a function of the light frequency $\Omega$ at 
zero temperature. The 
parameters are: $\Delta_{1,2}=\pm W$ and $\alpha_{L}=0.1$.
The lower curves correspond to a non-interacting system while the upper 
curves represent a dot coupled to LLs with $g=1/2$. 
}
\vspace*{-0.5cm}
\end{figure}
However, it is expected that the optical coupling is much smaller than that
to the contacting electrodes, 
so that we can expand (\ref{nonintTomega}) for
large $\alpha_{R,L}$. 
As a result we obtain the transmission coefficient in
form of two superimposed Lorentzians at energies $\Delta_{1,2}$ and widths
$\Gamma_{R,L}$. This implies for the $I-V$,
\begin{eqnarray}                    \label{pertIV}
 I(V) &=& \frac{G_0 2 \Gamma_R \Gamma_L W^2}{\pi} \\ \nonumber 
 &\times& \int_{-V/2}^{V/2} \frac{d
 \omega}{ [(\omega-\Delta_1)^2 + \Gamma_L^2][(\omega-\Delta_2)^2 +
 \Gamma_R^2]} \, .
\end{eqnarray}
The latter integral can easily be calculated in a closed form. 
The results are given the Appendix \ref{AppendixA}. 
In the case of symmetric electrode couplings, 
$\Gamma_R=\Gamma_L=\Gamma$, the current induced by the photon
absorption (we shall call it `ratchet' current) decays according to
$\sim W^2 \Gamma/\Omega^2$ at high frequencies. 
In the opposite infrared limit it varies linearly: 
\begin{eqnarray} \nonumber
 I \sim \frac{W^2 \Omega}{\Delta_1-\Delta_2} \sum_{i=1,2} \frac{\Delta_i}{
 \Delta_i^2+\Gamma^2} \, .
\end{eqnarray}
In the intermediate regime the ratchet current has its maximal 
value around $\Omega \sim \Delta_1-\Delta_2$, 
see Fig.\ref{Tzeroratchet}. 
A slight shift towards higher frequencies is a result of the mutual level
hybridisation due to tunnelling similar to that occurring in the case 
of a double dot. It turns out that coupling asymmetry does not 
result in any qualitative change in the induced current. 
The qualitative picture remains the same for all values 
of the optical coupling $W$. 
Depending on the sign of the external bias voltage the ratchet current 
enhances or suppresses the transport. This effect might be of immediate 
experimental relevance since it indicates the way the levels of the
quantum dot (or orbitals in the case of molecular dot) are arranged with
respect to contacting electrodes.  

As already discussed in Section \ref{noise} the knowledge of $T(\omega)$
enables not only to access the transport but also the zero--frequency 
noise properties using formula (\ref{zeroomeganoise}). For 
the calculation of the full frequency dependent noise spectrum the mere 
knowledge of $T(\omega)$ is insufficient because 
one needs the transmission 
amplitudes \cite{blanter}. In such a situation one can use the relation 
(\ref{Pviaspectrum}). Generally the optical coupling is expected to be 
relatively weak with respect to the lead-dot coupling so that we can calculate
the emission spectra at the leading order using Eqs.(\ref{matrixN1}):
\begin{eqnarray}    \label{sophint}
 &&S^E(\omega) = 2 \Gamma_R \Gamma_L \frac{W^2}{\pi} \Theta(V+\omega)
 \\ \nonumber
 &\times& \int_{-\omega-V/2}^{V/2} d \epsilon \, [(\epsilon-\Delta_1)^2 +
 \Gamma_L^2]^{-1} \nonumber [(\epsilon+\omega-\Delta_2)^2 + \Gamma_R^2]^{-1}
 \,. 
\end{eqnarray} 
This integral can be performed analytically but the result is 
lengthy. So we shall discuss only the special 
case of high bias voltage, corresponding to the FE via 
localised states. The evaluation of (\ref{sophint}) then yields
\begin{widetext}
\begin{eqnarray} \nonumber
 S^E(\omega) = 2 \Gamma_R \Gamma_L(\Gamma_R + \Gamma_L) \frac{W^2}{\pi}
 \frac{
 (\Delta_1+\Delta_2 + \omega)^2 + (\Gamma_L-\Gamma_R)^2}
 {[(\omega+\Delta_1-\Delta_2)^2 + \Gamma_L^2 +
 \Gamma_R^2]^2 - 2 \Gamma_L^2 \Gamma_R^2} \, ,  
\end{eqnarray} 
\end{widetext}
Since the emission process lowers the energy of the 
radiating system the actual spectrum is $\Theta(-\omega) S^E(\omega)$. 
Then this result describes a Lorentz-shaped peak around $\Delta_2-\Delta_1$, 
which is at odds with the experimental finding of Ref. \cite{french1}, 
where a superposition of two Gaussian peaks has been detected. 
There are three possible reasons for this
discrepancy, we list them in the order of relevance. First of all, the 
measuring apparatus can superimpose its own sensitivity curve, which is
usually of a Gaussian shape, over the actual luminescence spectrum. Secondly,
since the localised states on the nanotube tip could 
exist on the ends of free dangling $C-H$ bonds 
the finite temperature can 
lead to oscillations of the energy levels $\Delta_{1,2}$, which then could
become normally distributed. Another possible reason could be the electronic
correlations inside the nanotube. However, there is no \emph{a priori}
argument why their influence can result in Gaussian-shaped spectra.

\subsection{Interacting LL leads at $g=1/2$}            \label{Iratchet}
Now we turn to the case of interacting leads.
The Hamiltonian still contains the terms $H_K$, $H_C$, the 
new tunnelling contribution (\ref{newHt}) and the photon coupling term 
$H_W$ after the gauge transformation (\ref{gaugetrafo}). As usual, we 
apply the transformation (\ref{EKtrafo2}) and in order to access the Toulouse
point we again set the Coulomb coupling 
to $\lambda_C=2 \pi$. In the language of the new fermions defined by
(\ref{newfermions}) we obtain then the following 
Hamiltonian ($\Delta_{1,2}$ 
are assumed to be shifted by $\pm\Omega/2$): 
\begin{eqnarray} \nonumber
 H &=& H_0[\psi_\pm] + \sum_{i=1,2} \Delta_i d^\dag_i d_i \\ \nonumber
 &+& \Big[ \gamma_L \, d^\dag_1 \psi_-(0) + \gamma_R \, d^\dag_2 \psi^\dag_-(0)
 + W d^\dag_1 d_2 + \mbox{h.c.} \Big].
\end{eqnarray}
We proceed in the spirit of Section \ref{multiple} and derive the equations
of motion for the participating operators, 
\begin{eqnarray}                        \label{EoMsnew}
 (i\partial_t &-& \Delta_1) d_1 = W d_2 + \gamma_L \psi_-(0) \, , \nonumber \\
 (i\partial_t &-& \Delta_2) d_2 = W d_1 + \gamma_R \psi^\dag_-(0) \, , \\
 i[\psi^\dag_-(0^+) &-& \psi_-(0^-)] = \gamma_L d_1 + \gamma_R \psi^\dag_-(0)
 \nonumber \, , 
\end{eqnarray} 
where $0^\pm$ is again positive (negative) infinitesimal. Acting with $|Y|^2$
[see Eq.(\ref{xitrdef})] on the both sides of the third of Eqs.(\ref{EoMsnew})
we then use the first two ones in order to eliminate the dot operators. 
As a
result of this procedure we obtain an equation containing only $\psi_-$ 
operators, 
\begin{eqnarray}                     \label{newpsiequation}
 &&i|Y|^2[\psi^\dag_-(0^+) - \psi_-(0^-)] \\ \nonumber 
 &=& i Y^* [W \gamma_R \psi^\dag_-(0) +
 \gamma_L(i\partial_t - \Delta_2) \psi_-(0)] 
 \\ \nonumber 
 &+& \gamma_R Y [ W \gamma_L
 \psi^\dag_-(0) + \gamma_R (-i \partial_t - \Delta_2) \psi_-(0)] \, .
\end{eqnarray}
At this stage we again can make use of the decomposition
(\ref{partialdecomposition}) and reduce the relation (\ref{newpsiequation})
to exactly the form given in (\ref{finalequation}) with the same $E$ 
[see 
definition (\ref{Egammabetadefinitions})] but different $\gamma$ and 
$\beta_\pm$, 
\begin{eqnarray}
 \gamma &=& W \sqrt{\alpha_R \alpha_L} (\omega^2+\Delta_1 \Delta_2 - W^2) \, , 
 \nonumber \\
 \beta_+ &=& \Big\{ \alpha_R
 [(\omega+\Delta_1)(\omega+\Delta_2)-W^2](\omega-\Delta_1) \nonumber \\
 &-& \alpha_L
 [(\omega-\Delta_1)(\omega-\Delta_2)-W^2](\omega+\Delta_2) \Big\}/2 \, ,
 \nonumber \\ \nonumber
 \beta_-(\omega) &=& \beta_+(-\omega) \, ,
\end{eqnarray}
where we again normalised all energy variables to $\Gamma$ and defined 
$\alpha_{R,L} = \gamma^2_{R,L}/\Gamma$. 
With these conventions the total dot 
transmission coefficient is still 
given by the formula (\ref{1-D}) and the 
corresponding $I-V$ by Eq.(\ref{IVdefinition}). 

\begin{figure}[b]
\vspace*{0.5cm}
\includegraphics[scale=0.32]{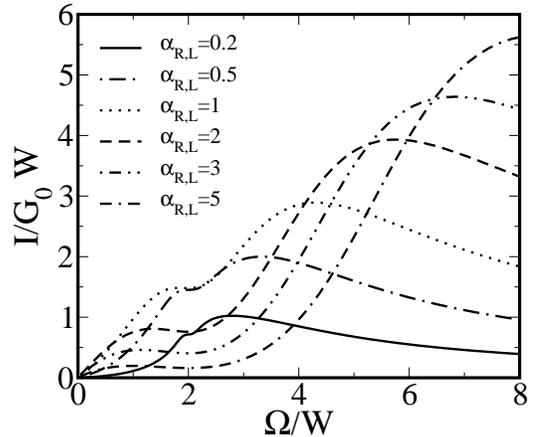}
\caption[]{\label{localmax} 
The same plot as in Fig.\ref{Tzeroratchet} 
for the symmetric interacting 
system for different lead-dot coupling strengths. 
}
\vspace*{-0.5cm}
\end{figure}

\begin{figure}[t]
\vspace*{0.5cm}
\includegraphics[scale=0.35]{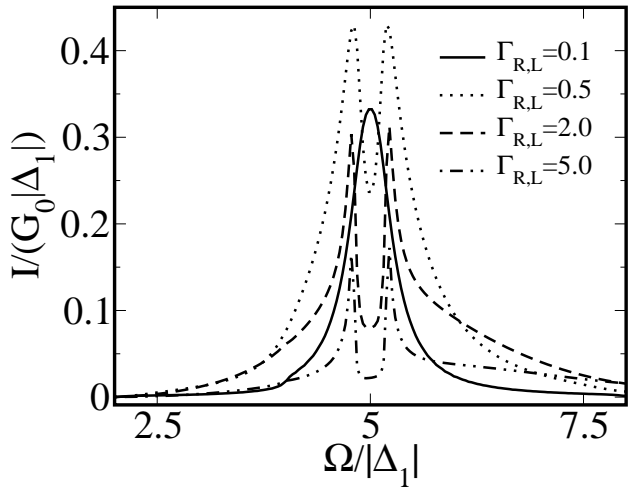}
\caption[]{\label{interacting} 
Ratchet current for a symmetric dot for different values of lead coupling.
All energies are measured in units of $|\Delta_1|$, $\Delta_1=-1$, 
$\Delta_2=4$ and $W=0.1$. 
}
\vspace*{-0.5cm}
\end{figure}

In the case of a weak dot-leads coupling and the symmetric level configuration
($\Delta_{1,2}= \pm \mbox{const}$), the ratchet current does not show
any significant change in comparison the the non-interacting case, see
Fig.\ref{Tzeroratchet}. The only difference is a slightly higher and 
wider maximum. However, as soon as the leads couple to the dot stronger 
than the dot levels among themselves an additional local minimum emerges
at low energies, see Fig.\ref{localmax}. 
This effect occurs only in the interacting system. More interesting features
arise in a system with weak optical coupling, see
Fig.\ref{interacting}. For positive $\Omega$ (which in our picture corresponds
to a situation favouring the ratchet effect) the distinguished peak at
$\Delta_1-\Delta_2$ which exists at $\Gamma\ll W$ splits in two
(which is actually a more relevant parameter range 
from the experimental point of view). This effect does not occur 
in the non-interacting situation. 
The origin of the total four peaks 
can be traced back to the presence of the LL 
zero-bias anomaly in vicinity of the Fermi energy. 
Concentrating only on the left
half of the system -- the left electrode with the level $\Delta_1$, one can
show in the lowest order in tunnelling that the correction to the level
spectral function (which at $\gamma_L=0$ is a delta function) 
is given by 
\cite{furusakimatveev}, 
\begin{eqnarray} \nonumber
 \delta A(\omega) \sim \gamma_L^2 \frac{\nu(\omega)}{(\omega-\Delta_1)^2} \, ,
\end{eqnarray}
where $\nu(\omega) \sim |\omega/\omega_c|^{1/g-1}$ is the DOS of a half-open
LL with a bandwidth $\omega_c$. At $g=1/2$ $\delta A(\omega)$ possesses two
maxima at $\pm \Delta_1$. This basic structure persists in all orders of
$\gamma_L$ with some corrections to the maxima positions. Exactly the same
thing happens in the right part of the system thus accounting for the 
total of four peaks seen in the ratchet current.

\section{Summary} \label{summary}

To conclude, we presented exact solutions of the 
single- and two-state 
non-interacting QDs coupled to interacting
(and non-interacting) electrodes. 
In both situations the corresponding Hamilton 
operators can be brought to a
quadratic form, and thus solved exactly, at the special interaction 
strength $g=1/2$ (the Toulouse point). 

In the first setup, the single-state QD, we recovered all results 
previously obtained via scattering formalism with the help of the 
non-equilibrium GF approach. 
We have shown, that exactly at the resonance such
model is fully equivalent to a simple tunnelling problem between two
interacting leads at a reciprocal interaction parameter $g=2$.
Furthermore, we derived equations for all possible 
GFs and applied them to calculate the noise properties of the system
which cannot be accessed by means of scattering formalism. 
Concentrating on the zero frequency noise properties we 
discussed the details of the voltage 
behaviour of the universal Fano factor $\nu_V$. 
It turns out to interpolate 
between the non-interacting value at high voltages 
and unity in the asymmetric
case (zero in the symmetric case) in the limit of small bias. In the 
asymmetric case $\nu_V$ possesses a 
minimum at some $V^*$, which is 
absent in non-interacting systems. 

In the case of many levels on the dot the question of the conductance
properties in the valley between the resonant tunnelling peaks is 
important. In order to shed light on this, we considered a model of 
a two-level QD, in the case when each of them is coupled to both leads. 
This additional feature does not 
destroy the solubility of the system at the Toulouse point and we succeeded
in applying the scattering formalism in this situation.  As expected,
the system turns out to show resonant transport signatures 
as long as the 
couplings are symmetric and one of the levels is tuned to the Fermi energies
in the electrodes. If the gate voltage tunes the system into the valley
between two peaks, $G(T)$ still keeps its on-resonance temperature dependence
with a renormalised pre-factor. 
However, turning on tunnelling between the two 
dot levels changes temperature dependence of the 
resonant transmission peaks 
in a non-trivial way. 

In the third model each level is assumed to be coupled only to 
one of the 
electrodes and the transport is supposed to be 
accompanied by absorption or
emission of photons. 
It turns out that the current through the system can 
flow even in the absence of any the bias voltage 
(ratchet current). In the 
non-interacting case the full non-linear $I-V$ 
as well as the absorption and 
emission spectra can be easily calculated via the Keldysh diagram approach.
We concentrated on the so called ratchet current induced solely by the
electromagnetic irradiation in absence of any voltage sources. As expected,
the dependence of the ratchet current on the light 
frequency possesses a 
clear maximum at the energy difference of the dot levels. As soon as we take
correlations into account, the picture changes considerably as the
peak splits in two with a pronounced minimum between them. 
The origin of 
this suppression can be traced
back to the zero--bias anomaly in the DOS of the interacting systems. 
Such spectacular effects make the ratchet current measurements 
an invaluable instrument for
studying interacting QD structures. In addition we derived an exact relation
between the absorption (emission) spectra and frequency dependent noise power
spectrum. It has important implications in the luminescence accompanying field
emission, which is believed to occur during cold electron emission from carbon
nanotubes. 

The key quantity, which generates an energy scale at which most of the 
predicted effects take place, is the lead--dot coupling $\Gamma$. In the
typical experiments made on semiconducting QDs $\Gamma$ ranges between $0.1-1$
$\mu$eV, which corresponds to temperatures around $1-10$ mK \cite{lundin}.
In the most current experiments conducted on contacted molecules the coupling
strength is expected to be even smaller \cite{c60,weber}.
For the results of Sections \ref{multiple} and \ref{noisefano} to be 
accessible in the experiments it is not necessary to go significantly below
the temperatures $T \sim \Gamma$. On the contrary, the experimental 
observability of all other predicted phenomena, and especially of the 
ratchet 
effects, depends crucially on the ability to either lower the temperatures
beyond the $T\sim \Gamma$ mark or build devices with high enough $\Gamma$
\footnote{Higher $\Gamma$ would automatically lead to higher
Kondo temperatures, which in QD can be as high as several K
\cite{goldhaber,cronenwett}. The Kondo physics would then obscure all
predicted effects. Therefore it is essential to keep the system spin-polarised
as assumed throughout the paper.}. 

\acknowledgements
The authors would like to thank H. Grabert for many valuable
discussions. This work was supported by the Landesstiftung Baden--W\"urttemberg
gGmbH (Germany), by the EC network DIENOW, and the EPSRC of the UK under
grants GR/N19359 and GR/R70309.

\section{Appendix A}      \label{AppendixA}
The result of integration in (\ref{pertIV}) is
\begin{widetext}
\begin{eqnarray} \nonumber
&& I(V) = G_0 \frac{2 \Gamma_R \Gamma_L W^2}{\pi} \\ \nonumber
&\times& \frac{\sum\limits_{i=1,2(L,R)}
 \Gamma_i [ (\Delta_1-\Delta_2)^2 -(-1)^i (\Gamma_R^2-\Gamma_L^2)]
 \sum\limits_{p=\pm}
 \tan^{-1} \left(\frac{V/2+p\Delta_i}{\Gamma_i}\right) +
 \Gamma_R \Gamma_L \sum\limits_{j=1,2(L,R)} p \Delta_i 
 \ln [\Gamma_j^2 + (\Delta_i
 -p(-1)^j V/2)^2]}{\Gamma_R \Gamma_L [ (\Delta_1-\Delta_2)^4 + 2
 (\Gamma_R^2+\Gamma_L^2)(\Delta_1-\Delta_2)^2 + (\Gamma_R^2-\Gamma_L^2)^2]} 
\end{eqnarray}
\end{widetext}

\bibliography{opt}
\end{document}